\documentclass{article}
\usepackage{amssymb, latexsym, amsmath}

\title{Exterior Differential Systems for Einstein Vacuum and Kaluza Gravity Theories}
\author{Frank B. Estabrook \\
Jet Propulsion Laboratory,  California Institute of Technology \\
4800 Oak Grove Drive,  Pasadena,  CA 91109}
\date{September 8, 2002}

\begin{document}
\maketitle

\begin{abstract}

We present  two families of exterior differential systems (EDS) for
causal embeddings of orthonormal frame bundles over Riemannian spaces of
dimension $q = 2,  3,  4,  5....$ into orthonormal frame bundles over flat spaces
of higher dimension.  We calculate Cartan characters showing that these
EDS are dynamical field theories.  The first family includes a
non-isometric  embedding  EDS  for classical Einstein vacuum relativity
($q = 4$). The second, generated only by 2-forms, is a family of classical
``stringy" or Kaluza-type ($q = 5$) integrable systems.  Cartan forms
are found for all these dynamical theories.
\end{abstract}

\section{Introduction}

        We discuss two families of geometric field theories.  The first
family is derived by variation of the Einstein-Hilbert action in $q =
3,  4,  5, $ etc. dimensions.  The case when $q = 4$ is vacuum general relativity
seen as a Ricci-flat orthonormal frame bundle over a 4-space which also
carries a second ``ghost"  metric induced from embedding in flat 10-space.  The
second family,  in dimensions  $q = 2,  3,  4,  5, $ etc., has a more
elegant string/gauge structure with $q-1-$dimensional frames that define
Riemannian submersions. When $q = 4$ or 5 it offers some alternatives
for classical field theory.

        By ``geometric" we mean that these theories are given as
exterior differential systems (EDS) for  embedding of q-dimensional
submanifolds ${R^q}$ in flat homogeneous isotropic metric spaces \({E^N}\) of higher
dimension,  say $N$. To formulate these EDS we in fact embed the
orthonormal frame bundles over the submanifolds into the orthonormal frame bundles over
the flat spaces,  that is,  into the groups ISO($N$), which have
dimension $N(N + 1)/2$. E. g., the $q = 4$ dimensional EDS are set using the 55 basis
1-forms of ISO(10).  The fibers of the embedded bundles are subgroups of
the O($N$)fibers of ISO($N$),  thus inducing embedding maps of their q-dimensional
bases ${R^q}$ into the ${E^N}$ bases of ISO($N$).

        By ``field theories" we mean that each of these various EDS is
shown,  by an explicit numerical calculation of its associated algebraic
structure (or sheaf),  to have the property of being ``causal".  Our technique for
analyzing the sheaf of an EDS,  using rational arithmetic and special
{\itshape Mathematica} programs,  is explained in Section 2.  In Cartan's theory of the
sequence of regular integral manifolds (of successively higher
dimensions,  using the sheaf) of a causal EDS,  the final construction of the embedding is
determined solely from boundary data,  and,  at least in the analytic
category, Cauchy existence and uniqueness are proved.  We believe that,  with
proper attention to signature,  the sets of partial differential
equations following from a causal EDS are those of a canonical field theory.  Non-trivial
embedding EDS that are causal are not common--we have been able to
discover just essentially these two families.  (There are also simple causal EDS
for embedding {\itshape flat} geodesic subspaces;  these are used in
Section 4).  A key to their existence may be that for all the EDS we consider
here we are also able to find Cartan $q$-forms from which the EDS may be
derived by arbitrary variation.

        There is a large literature,  beginning with Lepage and
Dedecker,  on the use of Cartan $q$-forms and their closure $q + 1$-forms
(``multisymplectic" forms).  These are the multidimensional field-theoretic extensions of
classical Hamiltonians and symplectic geometry.  A short but essential
bibliography can be found in Gotay \cite{got91};  cf. also Hermann \cite{her88} and
Estabrook \cite{est80}.  The differential geometric setting for that
work was for the most part (the structure equations of basis forms on) the first or second jet bundle
over a base of q independent variables.  Our use instead of basis forms
and structure equations from Lie rotation groups,  fibers of orthonormal
frame bundles over flat metric geometries,  appears to be an
innovation.  It allows application of those techniques of field theory to the movable frames of
general relativity,  and can lead to interesting extensions.

        In both these families of EDS we  generalize the method always
used in the mathematical literature for isometric embedding,  cf. e.g.
\cite{bry83},  \cite{gri87}, in that we do {\itshape not} require any alignment of the local frames
of the solution submanifolds with the orthonormal frames of the
embedding space.  Perhaps this generalization of the customary isometric
embedding  can be called ``dynamic embedding".  The EDS that naturally
arise are much more elegant.

        The Lie group ISO($N$),  or one of its signature siblings
ISO($N-1, 1$) etc.,  is the isometry group of $N$-dimensional flat space,
${E^N}$.  The group space is spanned by $N(N + 1)/2$ canonical vector fields,  and by a dual
basis of  left-invariant 1-forms that we denote by  ${{\theta }_{\mu }}, 
\mu = 1...N$,   corresponding to translations,  and by  ${{\omega
}_{{\mu \nu }}} = - {{\omega }_{{\nu \mu }}}$,   corresponding to rotations.  Now the structure equations for
general movable frames over an $N$-dimensional manifold are usually written
covariantly (on the second frame bundle) as
\begin{align}
{d\theta }^{\mu }+{\omega }^{\mu }_{\nu } \wedge {\theta }^{\nu } &= 0 \\
{d\omega }^{\mu }_{\nu }+{\omega }^{\mu }_{\sigma }\wedge {\omega
}^{\sigma }_{\nu }+{R^{\mu }}_{\nu
} &= 0.
\end{align}

These become the Cartan-Maurer equations of ISO($N$) or one of its
siblings when the curvature 2-forms \({{{R^{\mu }}}_{\nu }}\) are put
equal to zero, and upper indices are systematically lowered using (for signature) a
non-singular matrix of constants  \({{\eta }_{{\mu \nu }}}\),  after
which imposing antisymmetry (orthonormality)  ${{\omega }_{{\mu \nu
}}}\) = -  \({{\omega }_{{\nu \mu }}}$ .  These structure
equations then describe $N(N-1)/2$-dimensional rotation groups as fibers
over $N$-dimensional homogeneous spaces $E^N$.

 We will write the two families of EDS using partitions $(n,
m),  n + m = N$, of the basis forms of ISO($N$) into classes labeled
respectively by the first $n$ indices $i,  j$,  etc. $= 1,  2,  ... n$ and the
remaining indices $A,  B$,  etc. $= n +1,  n +2,  ... N$.  So the basis
forms are ${{\theta }_i},  {{\theta }_A},  {{\omega }_{{ij}}} = - {{\omega
}_{{ji}}},  {{\omega }_{{AB}}} = - {{\omega
}_{{BA}}},  {{\omega }_{{iA}}} =-{{\omega }_{{Ai}}}$.  Summation
conventions on repeated indices will be used separately on each partition.  
The structure equations (1) (2)become (for positive definite signature, 
${{\eta }_{{\mu \nu }}} = {{\delta }_{{\mu \nu }}}$)
\begin{align}
{d\theta }_{i}+{\omega }_{ij} \wedge {\theta }_{j} + {\omega }_{iA}
\wedge {\theta }_{A} &=0 \\
{d\theta }_{A}+{\omega }_{AB} \wedge {\theta }_{B} + {\omega }_{Ai}
\wedge {\theta }_{i} &=0 \\
{d\omega }_{ij}+{\omega }_{ik} \wedge {\theta }_{kj} - {\omega }_{iA}
\wedge
{\omega }_{jA} &= 0  \\
{d\omega }_{AB} +{\omega }_{AC} \wedge {\omega }_{CB} - {\omega }_{iA}
\wedge
{\omega }_{iB} &= 0  \\
{d\omega }_{iA} +{\omega }_{ij} \wedge {\omega }_{jA} + {\omega }_{AB}
\wedge
{\omega }_{iB} &= 0.
\end{align}

        It is a classic result \cite{gri87} that smooth local embedding
of Riemannian geometries of dimension $q = 3,  4,  5 ...$ is always
possible into flat spaces of dimension respectively $N =  6,  10,  15 ...$ ,
which motivates the partitions of our first family,  viz.  $(n, m) =  (3, 3),   (4, 6),
(5, 10), ...$ The causal EDS we give  determine submanifolds of ISO($N$) which
are themselves $O(n) \otimes O(m)$ bundles fibered over $q =
n$-dimensional base spaces,  say ${R^q}$ and induce maps of these into ${E^N}$. The $n$ 
${{\theta }_i}$  remain independent when pulled back to a solution
bundle,  satisfying the structure equations of an orthonormal basis in any cross
section,  and Equations ( 5) and ( 7) express embedding relations that
go back to Gauss and Codazzi. The solution bundle metric is the pullback
of ${{\theta }_i} {{\theta }_i}$.  We will present in Section 3 the
family of Einstein-Hilbert-Cartan forms from which the EDS of this first
family are derived by variation.  The EDS will require zero torsion for
the \({{\theta }_i}\) but  not insist on aligning the solutions with
these orthonormal frames (the \({{\theta }_A}\) are {\itshape not}
included in the EDS so it is not ``isometric",  and  from the induced map of
bundle bases there is also a less interesting ``ghost" metric which is
the pullback of ${{\theta }_i}{{\theta }_i}+{{\theta }_A}{{\theta }_A}$.)  The
induced curvature 2-forms are required by the EDS to satisfy
``horizontality" 3-form conditions and also to have vanishing Ricci $n-1$-forms.  We
calculate the Cartan characteristic integers describing the sheaves,
and showing the EDS to be well set and causal. These ``dynamic" embedding maps for low
values of N are shown in Figure  1.

        The field theories of our second family,  of dimension $q = 2,
3,  4,  5 ...$ also arise from embeddings into flat spaces  ${E^N}$ of
dimension $N = 3,  6,  10,  15,  ...$ but the EDS use different partitions,
viz..,  $(n, m) = (1, 2),  (2, 4),  (3, 7),  (4, 11)$,  etc.   Solutions
are ($O(n) \otimes O(m)$ bundles over) geometries  of dimension $q = n + 1$ 
and can be called n-branes.  They have rulings that are flat n-spaces.  The EDS
are generated only by sets of 2-forms (for vanishing torsion of both
partitions) and are so-called ``integrable systems".  Again
the embedding is dynamic,  the partitioned frames are not required to be aligned with
the solution manifolds.  In Section 4 we give the EDS and calculate the
Cartan characters showing them to be causal.   The  $n$ ${{\theta }_i}$ when
pulled back into a  solution determine a projection and  imply a metric
${{\theta }_i}{{\theta }_i}$ on its quotient space.  This is either  a theory of
relativistic rigidity or of a Kaluza-Klein gravitational field, depending
on $N$ and the signature adopted.   Cartan forms for those EDS are  easily
found.  The  embedding maps into ${E^N}$ are shown in Figure 2.

        As a sole illustration of introduction of coordinates into an
EDS,  the simplest of these non-isometric geometric field theories,
that  based on partition $(1, 2)$,  is integrated in Section 5.  Its solutions turn out
to be classically known,  in the guise of ruled surfaces in ${E^3}$.  We
have only changed signature to show it as a stringy field causally evolving
in time.

\section{Sheaf Algebra}

        We have previously expounded \cite{est89} Cartan's construction
of the so-called regular integral manifolds of a closed EDS,  as a
sequence,  or ``flag",  of simple  integrations (1) along a line,  beginning at a generic point,
(2) along a family of lines yielding a surface and containing the first
line, (3) along a two-parameter family of lines yielding a 3-space containing
the surface,  and so on.  This construction does not capture all
integral manifolds of the EDS,  and is in some disrepute mathematically,  as it
is of little use in proving new existence and uniqueness theorems for
p.d.e.'s, being limited to {\itshape analytic} solutions for subspaces.
Nonetheless Cartan's  theory shows the importance of a diagnostically
useful algebraic invariant of the EDS,  viz. the nested set of homogeneous linear
algebras or ``sheaf" that arises in the construction.  The sheaf
elegantly characterizes the analytic solutions of the EDS,  and above all can demonstrate
whether it is well set as a physical field theory.

        At a generic point in a space of dimension,  say,  $M$,  which
includes all independent,  dependent and gauge variables,   the sheaf of
a closed EDS is a nested set of linear homogeneous conditions on the $M$ components of a
set of  vectors ${V_1},  {V_2},  {V_3}, $ etc.  First a vector ${V_1}$
is found that annuls all the 1-forms in the EDS.  This condition is an
underdetermined set of linear equations for the components of ${V_1}$ so
a number of them,  say $M-{s_0}$,  may be chosen arbitrarily.  Given
that solution,  a new vector ${V_2}$ (independent of ${V_1}$) is
required that annuls all the 1-forms and also,  together with ${V_1}$,
annuls all the 2-forms in the EDS.  The rank of this second set of
linear equations,  for the components of ${V_2}$,  cannot be smaller than that
of the first set,  so we denote it ${s_0} + {s_1}$,
where ${s_1}$ is non-negative.  $M-{s_0} - {s_1}$ choices of remaining
components are made. Given the first two,  an independent vector
\({V_3}\)is then sought that annuls all 1-forms in the EDS,  that with either
${V_1}$ or ${V_2}$ annuls all 2-forms in the EDS,  and that with both
now also annuls all 3-forms in the EDS.  Again the rank cannot have
decreased,  so it is denoted ${s_0} + {s_1} + {s_2}$.  The $s$'s
are all non-negative.  The equations for new independent solution
vectors continue similarly,  and the sequence of Cartan characteristic
integers \({s_i}\) calculated.  At a generic point,  they are a set of numerical
concomitants of the EDS.  From them the causal and fiber structures of
the solutions,  integral submanifolds of the EDS,  can be inferred.

        It is straightforward \cite{est89} that the sheaf algebra
terminates:  there is a maximal generic number of solution vectors,  say
$g$.  This is Cartan's {\itshape genus,  }the dimension of the maximal regular 
solution submanifold.  In the following we will write the resulting set of characteristic sheaf
integers and the genus in the format $M\{{s_0}, {s_1}, {s_2}, ...{s_q},
...{s_{g-1}}\} g$.  The criterion for $g$ is that at the last step $M -
\sum _{i=0}^{g-1}{s_i} \geqslant g$.  Some arbitrary choices may
remain for \({V_g}, \) but further construction of independent
vectors fails at step $g + 1$.  ${V_g}$ is determined  (up to magnitude)
if there are no remaining choices for its components and the equality
holds; then,  at least in principle,  the solution of Cartan's construction is
uniquely related to suitable conditions set on its boundary.  We denote
such EDS as being ``causal".   $g$ will be the number of independent variables,
and $M-g$ the number of dependent variables in equivalent sets of
partial differential equations.

        It often happens that one of the integers (and consequently
those following it if there are no generating forms of those higher
ranks in the EDS), say ${s_q}$,  vanishes.  Then $q$ becomes the essential number of
independent variables in the resulting causal set of partial
differential equations, and $g-q$ variables in the solutions are gauge variables..
Geometrically,   there are in each solution $g-q-$dimensional fibers.
If $g-q$ variables are not explicitly present as basis forms in the EDS, their dual vectors
in the $M$-space are said to be Cartan characteristic vectors [7] of the
EDS,
and in that event the vanishing of the integer characters
${s_q}...{s_{g-1}}$ follows immediately.   For the genus $g$ we then
write ``$q +$ dim. fibers".

        To solve for the characteristic sheaf integers ${s_{i }}$
explicitly is not however easy,  and few sets are to be found in the
literature for nontrivial problems.  The coefficients in the ``linear" equations
for new vector components at each step depend on the components of
previous vectors,  and those components already have been required to jointly
annul the EDS at previous steps.  Consequently the components being
solved for at any step in fact are subject to both linear and non-linear constraint
equations,  which must be taken account of in determining the rank of
the new equations.  In general,  $p$-degree forms will lead to pth-order
polynomial constraints,  so if one attempts to obtain a solution for the
${s_i}$ algebraically this nested problem is not in fact a linear one.

        As an example from what follows,  consider an ideal in a space
of $M = 55$ dimensions,  so each $V$ has 55 components.  Calculating the
ranks of equations solving an EDS with  2-forms,  3-forms,  etc.,   when we come to
consider the equations for,  say,  ${V_6}$,  we would have as many as
100 quadratic constraints and 40 or so cubic constraints,  involving the 275
components of the previous five vectors ${V_1}...{V_5}$!   In a general
case it is unlikely that even a highly sophisticated algebraic manipulation
system would be capable of correctly evaluating the rank of the linear
equations for ${V_6}$ when subject to such constraints.

  This algebraic problem can be finessed by calculating particular
solutions with {\itshape numerical  }components.  At each step these
vector components are determined to satisfy the appropriate linear equations,  and any
components left undetermined are assigned {\itshape random} numerical
values, usually integers.  This purely numerical problem leading to a single
solution is now linear throughout,  and standard techniques to determine
the ranks of linear equations can be used,  even with a large number of
components.  The calculations are performed using rational arithmetic;
while not really essential,  this does avoid the possible problem of having
fortuitously small values become zero due to round-off errors.

        Clearly,  this technique suffers from the difficulty of any
Monte-Carlo approach in that it may not give the generic,  correct,
answer in any single calculation.  A particular set of random numerical components may  give
a lower rank than is true in general. This does in fact happen,
however very infrequently,  since the large number of random components
provides a very large ensemble.  Furthermore the calculation only
requires from a few seconds to a few minutes,  so it can easily be repeated
several times to check for accidental degeneracies.

\section{Einstein-Hilbert Action}

        The  EDS of our first family arise from Cartan $n$-forms on
ISO($N$) expressing the Ricci scalars of $q = n$ dimensional
submanifolds of ${E^N}$,
\begin{equation}
\Lambda  = {R_{{ij}}}\wedge {{\theta }_k}\wedge ...{{\theta }_p}
{{\epsilon }_{{ijk}...p}},
\end{equation}
where from the Gauss structure equation,  Eq. (5),  $2{R_{{ij}}} =
-{{\omega }_{{iA}}}\wedge \)\({{\omega }_{{jA}}}$
is the induced Riemann 2-form.   The exterior derivative of the $n$-form
field $\Lambda $ on ISO($N$),  using Eq. (3) and (7),  is quickly
calculated to be the $n + 1$-form (closed,  cosymplectic)
\begin{equation}
{d\Lambda } = {{\theta }_A}\wedge {{\omega }_{{Ai}}}\wedge
{R_{{jk}}}\wedge {{\theta
}_l}\wedge ...{{\theta }_p} {{\epsilon }_{{ijkl}...p}}.
\end{equation}

This $n+1$-form is a sum of products of the $m$ 1-forms ${{\theta
}_A}$ and the m $n$-forms ${{\omega }_{{Ai}}}\wedge \)\(
{R_{{jk}}}\wedge {{\theta }_l}\wedge \)...\({{\theta }_p}\) \({{\epsilon
}_{{ijkl}...p}}$.  A variational isometric embedding EDS is generated
by  the m  ${{\theta }_A}$,  their exterior derivatives for closure,  and the m
$n$-forms,  since any vector field that annuls it is sufficient to annul
${d\Lambda}$,  term by term.  That is,  up to boundary terms,  the arbitrary
variation of \(\Lambda \) vanishes on solutions.  We previously
calculated  Cartan's characteristic integers for these isometric embedding EDS  showing them
to be well set and causal \cite{est99}.  We denoted them as being
``constraint free" geometries.  Isometric embedding  formulations of the Ricci-flat field
equations then are obtained by adding in the closed $n-1$-forms for
Ricci-flatness as constraints. The augmented  EDS are again calculated to be causal.
Thus for vacuum general relativity,  the partition (4, 6),  the
constraint-free sheaf algebra was 55 $\{6, 6, 6, 12\} g = 4 + 21$ which,  with the
augumentation with four 3-forms became 55 $\{6, 6, 10, 8\} g  =  4  +
21$.  We now see that formulation as nevertheless somewhat unsatisfactory,  since
the Einstein-Hilbert action appears to have lead to equations which in
fact mostly follow from the imposed constraints.

        We have however noticed that there is another variational EDS
belonging to a different quadratic factoring of the cosymplectic forms
${d\Lambda }$,  Eq. (9).  The ${{\theta }_{i }}$ will frame a Riemannian metric on
an embedded space of dimension $n$ so long as the torsion 2-forms
${{\omega }_{{iA}}}\wedge {{\theta }_A}$ vanish,  and these factor Eq.
(9) term-by-term,  as products with the n $n-1$-forms for
Ricci-flatness. The exterior derivatives of the torsion terms are 3-forms for
symmetries of the Riemann tensor,  sometimes called conditions for
horizontality. In sum,  we have considered the following closed EDS (which now do
{\itshape not} include the mathematically customary isometric embedding
1-forms ${{\theta }_A}$)
\begin{equation}
({{\omega }_{{iA}}}\wedge {{\theta }_A},  {R_{{ij}}}\wedge {{\theta
}_j}, {R_{{ij}}}\wedge
{{\theta }_k}\wedge ...{{\theta }_l}{{\epsilon }_{{ijk}...{lp}}})
\end{equation}

The sheaf calculation shows these to be  well set and causal systems for
embedding of $O(n) \otimes O(m)$ bundles over $n$ space,  for the
partitions (3, 3),  (4, 6),  (5, 10) etc. as stated in the introduction.  The
embedding dimension,  the computed Cartan characters,  genus and
$O(n) \otimes O(m)$ fiber dimension of the solutions for these cases are 
respectively $21\{0, 6, 3\} g = 3 + 9,  55\{0, 4, 12, 14\} g = 4  +
21,   120\{0, 5, 10, 20, 25\} g = 5 + 55$,  etc.  These causal dynamic 
embeddings are shown in Figure 1.

 The base spaces of the fibered solution manifolds are spanned by the
1-forms ${{\theta }_i}$; evidently a solution is a bundle of orthonormal
frames belonging to the Ricci-flat Riemannian connection ${{\omega }_{{ij}}}$, together with
a gauge connection ${{\omega }_{{AB}}}$.  The metric is ${{\theta
}_i}{{\theta }_i}$. There is also present in the base space ${R^n}$ 
another metric pulled back from the induced embedding
of it in the base space ${E^N}$ about which we know little.   It
is a ghost tensor field,  perhaps with only indirect influence.  The
ideals we are writing are set on ISO($N$),  and their solutions are frame
bundles embedded in ISO($N$),  and the induced embeddings of the base
spaces seem to be of less interest.

         The ideal Eq. (10) is contained in the augmented isometric
embedding ideal we have previously used,  so solutions of the latter
will be solutions of the former.  This would seem to imply that our new dynamic embedding
ideal will have additional solutions; indeed it implies fewer partial
differential equations than does the isometric embedding ideal augmented with
constraints for Ricci-flat geometry.  Perhaps so-called singular
solutions of the isometric embedding ideal--solutions which are not regular,  that is,
obtained by Cartan's sequential integrations--now appear as regular
solutions, which could make this new formulation important for local numerical
computation from boundaries.

\section{Torsion-free  n-brane Embedding}

 We have searched whether the torsion 2-forms induced in {\itshape both}
the local partitions can together be taken as an EDS:
(${{\omega }_{{iA}}}\wedge
{{\theta }_A}\),  \({{\omega }_{{iA}}}\wedge {{\theta }_i}$).
%({{\theta }_i}{{\theta
%}_i}+{{\theta }_A}{{\theta }_A},  {{\omega }_{{iA}}}\wedge
%{{\theta }_A}).
It can easily be checked that it is closed,  and calculation
of the characteristic integers of the sheaf indeed showed that
{\itshape for just the values of (n, m) of the second family described in the
introduction} these EDS are  causal,  with $q = n + 1$ and fibers $O(n) \otimes
O(m)$, dim $n(n-1)/2 + m(m-1)/2$.  The results for the first five EDS
are:
$(n, m) = (1, 2)$,  $6\{0, 3\} q = 2 + 1$ dim fiber; $(2, 4)$, $21\{0, 6, 5\}
3 + 7$; $(3, 7)$, $55\{0, 10, 9, 8\}4$ + 24;  $(4, 11)$, $120\{0, 15, 14, 13,
12\} 5$ + 61;  $(5, 16)$, $231\{0, 21, 20, 19, 18, 17]6$ + 130; and the pattern
seems evident.  The embedding maps are shown in Figure 2.

   Well set EDS for geodesic flat dimension n submanifolds of flat
N spaces are generated,  using the partition $(n, m)$,  by the closed ideal
of 1-forms $({{\theta }_{A }}, {{\omega }_{{Ai}}}$).
For example,  if $N = 3$ and $n = 1$ and $m = 2$,   geodesic lines in
flat 3-space,  the Cartan characteristic integers are
$6\{4\}1 +1$.  If $N = 4$,  for partition $(1, 3)$ we find $10\{6\}1 +
3$ (in all cases ${{\omega }_{{ij}}}$ and
${\omega }_{{AB}}$
%${{\theta }_{A }}, {{\omega }_{{Ai}}}$ and
%${{\omega }_{{ij} }}$
give the Cauchy characteristic fibers).  Similarly,  the EDS for flat
2-dimensional submanifolds of flat $N$ spaces are generated by the 1-forms 
with partitions $(2, N-2)$.  For example if $N =  5$, $(n, m) = (2, 3)$,  and  the sheaf
algebra is $15\{9, 0\}2 + 4$. When $N = 6,  (n, m) = (2, 4)$ and $21\{12, 0\}2 + 7$.  These
constructions clearly continue.  Now the torsion-free EDS
(${{\omega }_{{iA}}}\wedge {{\theta }_A}$,  ${{\omega }_{{Ai}}}\wedge
{{\theta }_i}$) is contained in (${{\theta }_{A }}, {{\omega
}_{{Ai}}}$), so we see that the q-dimensional solutions of the torsion-free
embedding theory must contain flat geodesic fibers of dimension $n = q-1$.  Thus 
the solutions are {\itshape ruled }spaces.

        In a solution the \({{\theta }_{i }}\) remain independent (are
``in involution") but fall short by one of being a complete basis.  They
define there a vector  field,  say {\itshape V},  of arbitrary normalization (a
congruence),  by the relations   \(V\cdot  {{\theta }_i}\){\itshape
 } \(=\)  \(V\cdot  {{\omega }_{{ij}}}\) \(=\) {\itshape  }\(V\cdot
{{\omega }_{{AB}}}\)\( \)\(=\)  0.
 Contracting {\itshape V} on the second torsion 2-form,  since the 
${{\theta }_i}$ remain linearly independent,  gives also $V\cdot
{\omega }_{iA}= 0$.  Thus the Lie derivative with respect to V of all of
these (except for \({{\theta }_A}\)) vanishes.  They live in an
n dimensional quotient space of the solution,  with metric \({{\theta
}_i}{{\theta }_i} \) and Riemann tensor \({{\omega }_{{iA}}}\wedge
{{\omega }_{{Aj}}} \).  This is in contrast to the rulings,
n-dimensional subspaces also carrying the metric  \({{\theta
}_i}{{\theta }_i}\) but flat.

        In an earlier time we have discussed the problem of defining a
rigid body in special and general relativity \cite{wah66}.  The
kinematic quotient-space definition of rigidity due originally to Max Born (
Riemannian submersion [2]) was shown by Herglotz and Noether to have only 
three degrees  of freedom: the only Born-rigid congruences which were rotating (
had vorticity) in Minkowski space were isometries of the space-time without time
evolution.  We showed this to be the case also for kinematic or ``test" rigid bodies
moving in vacuum Einstein spaces.   It seemed to be impossible then to
sensibly discuss the so-called ``dynamic" rigid bodies introduced by Pirani,  which
were to carry their own three dimensional geometry. We are charmed by having
now arrived at  space-times,  using dynamic embedding in the (3, 7) partition,
having the greater dynamical freedom allowed by separation of the roles
of the induced metrics in the cross sections and quotient space of a solution.

        In the (4,  11) partition,  the solutions are five dimensional,
with a dynamically rigid congruence that projects to a metric
4-space.  This is
surely a well-posed causal variant of Kaluza-Klein theory,  and remains
for further investigation.

        Closed EDS generated only by ``invariant" 2-forms  (meaning no
explicit functional dependence,  as here) have a special structure,
inasmuch as they
can be equivalent to dual infinite Lie algebras of Kac-Moody type and
lead to hierarchies of so-called integrable systems.  The prototype of
this is the well-known Korteweg-de Vries equation,  which both leads to
\cite{est90},  and belongs to the hierarchy of,  the infinite Lie
algebra \({{{A_1}}^{(1)}}\)derived from SL(2, R).   The Kac-Moody algebras 
dual to our embedding EDS remain to be worked out.

        Although we did not derive these EDS variationally,  Cartan
forms are easily found.  For example,  in the (3, 7) theory either the
2-forms \({{\tau }_{A }} = {{\omega }_{{Ai}}}\wedge {{\theta }_i}\) or \({{\sigma }_i} =
{{\omega }_{{iA}}}\wedge {{\theta }_{A }}\) can be used to write a quadratic 
Cartan form as in some Yang-Mills theories:
\begin{equation}
\Lambda  = {{\tau }_A}\wedge {{\tau }_A},
\hspace{0.05cm}\mbox{so}\hspace{0.2cm} d{\Lambda } =
2 {{\tau }_A}\wedge {{\omega }_{{Ai}}}\wedge {{\sigma }_i}
\end{equation}

Every term of $d\Lambda$  contains both a ${{\tau }_A}$ and a ${{\sigma }_i}$
so arbitrary variation yields the EDS.  We also note that
${{\tau }_A}\wedge {{\tau }_A} + {{\sigma }_i}\wedge {{\sigma
}_i}$ is exact.

\section{The Partition (1, 2)}

        We will set this EDS on the frame bundle ISO(1, 2) over a flat
3-space with signature (-, \(+\), -),  so the structure equations of the
basis are
\begin{align}
{d\theta }_{1} + {\omega }_{12} \wedge {\theta }_{2} + {\omega }_{31}
\wedge {\theta }_{3} &= 0  \\
{d\theta }_{2} + {\omega }_{12} \wedge {\theta }_{1} - {\omega }_{23}
\wedge {\theta }_{3} &= 0  \\
{d\theta }_{3} - {\omega }_{31} \wedge {\theta }_{1} - {\omega }_{23}
\wedge {\theta }_{2} &= 0  \\
{d\omega }_{12} - {\omega }_{31} \wedge {\omega }_{23} &= 0  \\
{d\omega }_{23} - {\omega }_{12} \wedge {\omega }_{31} &= 0  \\
{d\omega }_{31} + {\omega }_{23} \wedge {\omega }_{12} &= 0,
\end{align}
and the EDS to be integrated is generated by the three 2-forms ${\omega
}_{iA} \wedge {\theta }^{A},   {\omega }_{Ai} \wedge
{\theta }^{i},   i = 1,   A = 2, 3$:
\begin{align}
{\omega }_{12} \wedge {\theta }_{2} + {\omega }_{31} &\wedge {\theta
}_{3}  \\
{\omega }_{12} &\wedge {\theta }_{1}  \\
{\omega }_{31} &\wedge {\theta }_{1}.
\end{align}

The characteristic integers of the sheaf are 6\{0, 3\} g \(=\) 2 with
O(2) fiber (since ${\omega }_{23}$ is not present). To introduce
coordinates - scalar fields - we will successively prolong the EDS with potentials
or pseudopotentials,  checking at each step that it remains well-set and
causal.

        First,  it is obvious that there is a conservation law,  a
closed 2-form that is zero mod the EDS,  viz. \({{{d\theta }}_1}\). So
we adjoin the 1-form
\begin{equation}
{{\theta }_1}+{dv},
\end{equation}
introducing the scalar potential $v$. The sheaf now is $7\{1, 3\} g = 2$
with O(2) fiber. Next we specialize to a particular,  convenient,  fiber
cross-section making a choice of frame or gauge: we introduce two new fields \(\zeta
\) and  \(\eta \) while prolonging with three 1-forms taken so that the
original 2-forms in the EDS vanish (they have been ``factored")
\begin{align}
{{\omega }_{12}}-\zeta  {{\theta }_1}  \\
{{\omega }_{13}}-\eta  {{\theta }_1}\zeta   \\
\zeta  {{\theta }_2}-\eta  {{\theta }_3}+(\eta +\zeta ){{\theta }_1}.
\end{align}

To maintain closure,  however,  three new 2-forms,  exterior derivatives
of these or algebraically equivalent,  must also be adjoined:
\begin{align}
(d{\zeta } - \eta {\omega }_{23}) &\wedge {dv}  \\
(d{\eta } - \zeta {\omega }_{23}) &\wedge {dv}  \\
(\eta {d\zeta }- \zeta {d\eta }) \wedge ({{\theta }_2}+{{\theta }_3})-
(\eta + \zeta ){\omega }_{23} &\wedge (\eta {\theta }_{2} - \zeta
{\theta }_{3}).
\end{align}

Now we have $9\{4, 3\} g = 2$ with no remaining gauge freedom. ${\omega
}_{23}$ now appears in the EDS,  but is conserved,  ${d\omega
}_{23}= 0$ mod EDS. Thus,  we can introduce a pseudopotential variable
$x$ ,  and then further find another conserved 1-form and a final
pseudopotential $u$. Which is to say we can adjoin
\begin{align}
{{\omega }_{23}}-{dx}  \\
{{\theta }_2}+{{\theta }_3}-{e^x}{du},
\end{align}
without adding any 2-forms to the EDS. We have a total of 11 basis
1-forms: six in ${\theta }_{i}$,  ${\theta }_{A}$,  ${\omega }_{AB}$,
${\omega }_{iA}$,  plus $d\zeta,  d\eta,  dx,  du,  dv$,  and an EDS
with $11\{6, 3\} g = 2$. The pulled-back original six bases
are now all solvable in terms of coordinate fields on the solutions,
and can be eliminated: $5\{0, 3\} g = 2$.  This is equivalent to a set
of first order partial differential equations in 3 dependent variables and 2
independent variables.

        Taking $x$ and $v$ as independent in the solution,  we can solve
the first two 2-forms in Eq. (25) and (26) for \(\eta \) and \(\zeta \):
\begin{align}
\eta  &= a {e^x}+b {e^{-x}}  \\
\zeta &=a {e^x}-b {e^{-x}},
\end{align}
where $a$ and $b$ are arbitrary functions of $v$. The third 2-form then
amounts to
\begin{align}
{e^x}&= 1/2{{(b/a)}^\prime} {{\partial }_x}u,  \\
\mbox{which integrates to}\hspace{0.3cm} {e^x}&= 1/2{B^\prime}(u-A(v)).
\end{align}
We have put $b/a = B(v)$ and prime is derivation with respect to $v$.

        The two arbitrary functions of $v$,  $A$ and $B$,  give the
general solution. On it the pulled-back bases of ${E^3}$ (no longer
orthonormal or independent)are
\begin{align}
{{\theta }_1} &= - {dv}  \\
{{\theta }_2} &= {dv} +\frac{a {e^x}+b {e^{-x}}}{2 a}{du}  \\
{{\theta }_3} &= - {dv} +\frac{a {e^x}-b {e^{-x}}}{2 a}{du},
\end{align}
and the induced  2-metric from \({E^3}\) is
\begin{align}
g &= - {{\theta }_1}{{\theta }_1}+{{\theta }_2}{{\theta }_2}-{{\theta
}_3}{{\theta }_3}  \\
&= B {{{du}}^2}+{B^ \prime }(u-A){dv} {du}- {{{dv}}^2}.\notag
\end{align}

This is,  up to signature,  the metric found classically
from the construction of ruled surfaces in \({E^3}\),  cf, e.g.,
Eisenhardt
[10].  The  surfaces are intrinsically characterized by a ``line of
striction",  the locus $u-A(v) = 0$,  and a ``parameter of
distribution"  $2B/{B^{\prime
}}$.  The  geodesic rulings,  on which ${{\theta }_2}$,  ${{\theta }_3}$,
${{\omega }_{12}}$,  and ${{\omega }_{13}}$ pull back
to vanish,  are the set of lines $u =$ const.  The rigid congruence is
the set of lines on which V contracted with ${\theta }_{1}$,  ${\omega
}_{12}$ and ${\omega }_{13}$ vanishes,  hence $v =$ const.

\subsection*{Acknowledgements}

This work has at every point benefitted from conversations with my long
time colleague Hugo D. Wahlquist.  It follows on several lines of
research we have jointly pursued.  Wahlquist first suggested the non-isometric
embedding EDS for vacuum relativity ($q = 4$).  The Cartan characters
reported were all obtained using his suite of {\itshape Mathematica} programs for
manipulation of algebra-valued-forms and calculation of the ranks of
sheaf algebras.  This research was performed at the Jet Propulsion
Laboratory,  California Institute of Technology,  under contract with
the National Aeronautics and Space Administration.

\subsection*{Figure Captions}

Figure 1.  Einstein-Hilbert field theories for embedding partitions $(n,
m),  N = n + m$.  The causal sheaf algebra is tabulated in each case
as dim ISO($N$) $\{{\omega }_{13},  {s_0},  ... {s_1}\} q +$ dim
fiber.

Figure 2.  Torsion-free field theories for embedding partitions $(n,
m),   n + m = N$.  Sheaf notation as in Fig. 1.

\end{document}